\documentclass[11pt,twoside,a4paper,cmspaper,final,collab]{cms-tdr}
\def\svnVersion{d526bf6-D}\def\svnDate{2026/11/09}\def\cmsCernNoTag{CERN-EP-2026-291}\def\cmsCernDate{\today}\def\cmsMessage{Submitted to Physical Review Letters}
\begin{document}\cmsNoteHeader{BPH-26-004}

\newcommand{\twomu}{\ensuremath{\PGh\to\PGmp\PGmm}\xspace}
\newcommand{\twomutwopi}{\ensuremath{\PGh\to\PGmp\PGmm\PGpp\PGpm}\xspace}
\newcommand{\fourmu}{\ensuremath{\PGh\to\PGmp\PGmm\PGmp\PGmm}\xspace}
\newcommand{\twomutwoe}{\ensuremath{\PGh\to\PGmp\PGmm\Pep\Pem}\xspace}

\newcommand{\Ntwomu}{\ensuremath{N_{2\PGm}^{i,j}}\xspace}
\newcommand{\Btwomu}{\ensuremath{\mathcal{B}_{2\PGm}}\xspace}
\newcommand{\Bfourmu}{\ensuremath{\mathcal{B}_{4\PGm}}\xspace}
\newcommand{\Btwomutwopi}{\ensuremath{\mathcal{B}_{2\PGm 2\PGp}}\xspace}
\newcommand{\Btwomutwoe}{\ensuremath{\mathcal{B}_{2\PGm 2\Pe}}\xspace}
\newcommand{\mtwomu}{\ensuremath{m_{2\PGm}}\xspace}
\newcommand{\mfourmu}{\ensuremath{m_{4\PGm}}\xspace}
\newcommand{\mtwomutwopi}{\ensuremath{m_{2\PGm 2\PGp}}\xspace}
\newcommand{\twomugamma}{\ensuremath{\PGh\to\PGmp\PGmm\PGg}\xspace}
\newcommand{\twopigamma}{\ensuremath{\PGh\to\PGpp\PGpm\PGg}\xspace}
\newcommand{\twopipizero}{\ensuremath{\PGh\to\PGpp\PGpm\PGpz}\xspace}
\newcommand{\meta}{\ensuremath{m_{\PGh}}\xspace}
\newcommand{\gammatotwomu}{\ensuremath{\PGg\to\PGmp\PGmm}\xspace}
\newcommand{\bkgmodela}{\ensuremath{(\meta - m_0)^{p_0}}\xspace}
\newcommand{\bkgmodelb}{\ensuremath{(\meta-m_0)^{p_0} \exp(p_1 \meta)}\xspace}
\newcommand{\bkgmodelc}{\ensuremath{(\meta-m_0)^{p_0} (1 + p_1 \meta)}\xspace}
\newlength\cmsFigWidth
\ifthenelse{\boolean{cms@external}}
{
  \setlength{\cmsFigWidth}{0.49\textwidth}
}
{
  \setlength{\cmsFigWidth}{0.6\textwidth}
}

\cmsNoteHeader{BPH-26-004}
\title{Observation of the \texorpdfstring{$\PGh\to\PGmp\PGmm\PGpp\PGpm$}{eta to mu+mu-pi+pi-} decay and an improved measurement of the \texorpdfstring{$\PGh\to\PGmp\PGmm\PGmp\PGmm$}{eta to mu+mu-mu+mu-} branching fraction}

\author[cern]{The CMS Collaboration}

\date{\today}

\abstract{An observation of the rare decay \twomutwopi is reported from proton-proton collision data at a center-of-mass energy of 13.6\TeV, collected by the CMS experiment at the CERN LHC in 2022 and 2023, and corresponding to an integrated luminosity of 62.6\fbinv. In addition, the branching fraction ratio of \fourmu relative to \twomu is measured with a precision improved by 44\% compared with the previous measurement. The measured branching fraction ratios are $\mathcal{B}(\twomutwopi)/\mathcal{B}(\twomu) = [2.89 \pm 0.48\stat{}^{+0.42}_{-0.29}\syst]\times10^{-3}$ and $\mathcal{B}(\fourmu)/\mathcal{B}(\twomu) = [0.86 \pm 0.06\stat{}^{+0.09}_{-0.08}\syst]\times10^{-3}$.}

\hypersetup{
pdfauthor={CMS Collaboration},
pdftitle={Observation of the eta to mu+mu-pi+pi- decay and an improved measurement of the  eta to mu+mu-mu+mu- branching fraction},
pdfsubject={CMS},
pdfkeywords={CMS, eta meson, rare decays, double-Dalitz decay, eta to mu+mu-pi+pi-, eta to mu+mu-mu+mu-, branching fraction, transition form factor, chiral perturbation theory, data parking}}

\maketitle

Rare decays of the \PGh meson provide sensitive probes of low-energy quantum chromodynamics (QCD) and possible contributions from physics beyond the standard model (BSM).
These decays are described within chiral perturbation theory (\ensuremath{\chi}\text{PT})~\cite{Bijnens:2002qy} and its extensions, such as vector meson dominance (VMD)~\cite{10.1143/PTP.73.926,Bando:1985rf}, as well as by phenomenological approaches incorporating experimental data~\cite{escribanoDatadrivenApproachPi2018}.
They also provide input to the dispersive determination of the \PGh meson transition form factor, in particular its doubly virtual behavior, which governs the hadronic light-by-light contribution to the anomalous magnetic moment of the muon~\cite{Gan:2020aco}.
Deviations from theoretical predictions could indicate contributions from BSM weakly coupled particles below the GeV scale, including dark photons, hidden gauge bosons, light Higgs scalars, and axion-like particles~\cite{Gan:2020aco}.

The \PGh meson is produced copiously in proton-proton ($\Pp\Pp$) collisions at the CERN LHC, but its low mass~\cite{PDG2026} leads to highly collimated decay products, making their reconstruction experimentally challenging.
Purely leptonic double-Dalitz decays of the \PGh meson, such as \fourmu and
\twomutwoe, proceed via internal conversion of two virtual photons and are
strongly suppressed, with the larger mass of the muon compared with that of
the electron further reducing the available phase space for \fourmu relative
to \twomutwoe.
Their branching fractions have been measured by the CMS Collaboration to be $\Bfourmu=(5.0 \pm 1.3)\times10^{-9}$ and $\Btwomutwoe=(2.4 \pm 0.8)\times10^{-6}$~\cite{cms4mu,CMS-PAS-BPH-24-001,PDG2026}.
These measurements are consistent with the predicted branching fraction for \fourmu of $(3.98 \pm
0.15)\times10^{-9}$~\cite{escribanoDatadrivenApproachPi2018} and the latest predictions for
\twomutwoe in the range of $(2.3\text{--}2.4)\times10^{-6}$ based on data-driven approaches~\cite{Kampf,escribanoDatadrivenApproachPi2018}.

In contrast, the mixed hadronic-leptonic decays, such as \twomutwopi, probe both electromagnetic and strong-interaction dynamics through an intermediate virtual photon, with possible contributions from the QCD box anomaly~\cite{Gan:2020aco}. 
Theoretical predictions for \twomutwopi are in the $(6\text{--}12)\times10^{-9}$ range, depending on the treatment of the relevant transition form factor and decay dynamics~\cite{Faessler:1999de,borasoyEtaEtaPrime2007,Escribano_2015,Zillinger:2022eva}.
This decay has not yet been observed. The most stringent existing limit is $\Btwomutwopi < 4.0\times10^{-7}$ at 90\% confidence level (\CL), reported by the BESIII Collaboration~\cite{BESIII:2025cky}.
Measurements of these rare decays test theoretical descriptions of the \PGh meson structure in a kinematic regime with limited prior experimental constraints.

This Letter reports the observation of the rare decay \twomutwopi and an improved measurement of the \fourmu branching fraction.
The $\Pp\Pp$ collision data were collected by the CMS experiment at the LHC at a center-of-mass energy of 13.6\TeV in 2022 and 2023, corresponding to an integrated luminosity of 62.6\fbinv.
The branching fractions are measured relative to that of the normalization channel \twomu,
\begin{equation}\label{eq:br}
\frac{\mathcal{B}_{X}}{\Btwomu}
= \frac{N_{X}}
{ \sum_{i,j}
  N_{2\PGm}^{i,j}\,
  \frac{\epsilon_{X}^{i,j}}{\epsilon_{2\PGm}^{i,j}} },
\end{equation}
where $X$ denotes the signal channel (\twomutwopi or \fourmu), $N_{X}$ is the corresponding observed
signal yield, and $N_{2\PGm}^{i,j}$ is the number of observed \twomu events in the $(i, j)$ bin,
where $i$ and $j$ index the \PGh meson transverse momentum (\pt) and absolute rapidity $(\abs{y})$ bins, respectively. 
The quantities $\epsilon_{X}^{i,j}$ and $\epsilon_{2\PGm}^{i,j}$ are the corresponding total efficiencies in the $(i, j)$ bin for channels $X$ and \twomu, respectively.
Tabulated results are provided in the HEPData record for this analysis~\cite{hepdata}.

The CMS apparatus~\cite{CMS:2008xjf,CMS:2023gfb} is a multipurpose detector designed to trigger on and identify electrons, muons, photons, and (charged and neutral)
hadrons~\cite{CMS:2020uim,CMS:2018rym,CMS:2014pgm}. A superconducting solenoid of 6\unit{m} internal diameter provides a magnetic
field of 3.8\unit{T}. Within the solenoid volume are the silicon pixel and strip tracker, a crystal electromagnetic calorimeter,
and a brass and scintillator hadron calorimeter. Muons are measured in gas-ionization detectors embedded in the steel flux-return
yoke outside the solenoid, covering the pseudorapidity range $\abs{\eta} < 2.4$. A more detailed description of the CMS detector,
together with a definition of the coordinate system used and the relevant kinematic variables, can be found in Refs.~\cite{CMS:2008xjf,CMS:2023gfb}.

Events of interest are selected using a two-tiered trigger system. The first level, composed of custom hardware processors, uses information from the calorimeters and muon detectors to select events at a rate of around 100\unit{kHz} within a fixed latency of 4\mus~\cite{CMS:2020cmk}. The high-level trigger consists of a farm of processors running a version of the full event reconstruction software optimized for fast processing, and reduces the event rate to about 5\unit{kHz} before data storage~\cite{CMS:2016ngn,CMS:2024aqx}.
The trigger used to collect the signal events for this analysis required a pair of opposite-sign muons, 
with one muon having $\pt>4\GeV$ and the other $\pt>3\GeV$. 
This trigger was part of a data-parking framework~\cite{parkingtrigger}, introduced in 2022, 
in which events selected by inclusive trigger algorithms with low \pt threshold were saved for later reconstruction when resources were available.

Simulated samples are used to model the \twomu, \twomutwopi, and \fourmu decays. They are generated using the \textsc{Pluto} framework~\cite{Frohlich:2007bi}, with the transition form factors of the virtual photons described by the VMD model~\cite{SAKURAI19601,Faessler:1999de,Escribano_2015,PhysRevD.94.054033}.
The simulated \twomutwopi events are reweighted to the dimuon and dipion invariant mass distributions predicted in the chiral unitary approach~\cite{borasoyEtaEtaPrime2007}. The effect of the theoretical uncertainty in these distributions on the efficiency is negligible.
Each generated \PGh meson is boosted to the laboratory frame with \pt and rapidity uniformly distributed in the 5--70\GeV and $\abs{y}<2.4$ ranges, covering the kinematic region used for the analysis, following the procedure reported in previous CMS publications~\cite{cms4mu,CMS-PAS-BPH-24-001}.
To evaluate the candidate selection efficiency in events with more than one \twomutwopi candidate, additional simulated \twomutwopi events are generated with the SoftQCD:nonDiffractive option in \PYTHIA 8.3~\cite{Bierlich:2022pfr}, using the CP5 underlying event tune~\cite{CMS:2019csb}, and decays of the \PGh meson are simulated using \EVTGEN 1.3.0~\cite{Lange:2001uf}. The decay products from all simulated events are propagated through a detailed \GEANTfour-based simulation~\cite{Agostinelli:2002hh} of the CMS detector and additional $\Pp\Pp$ interactions per bunch crossing (pileup) are added to the simulated samples with a multiplicity distribution matching the measured one.

Muon candidates matched to the trigger muons must pass the medium identification criteria~\cite{CMS:2018rym}, and at least two such muons with opposite charges are required in each event for the two signal channels and the normalization channel. The leading (subleading) muon must satisfy $\pt>4\,(3)\GeV$, and the vertex fit of the muon pair must have a probability greater than 0.5\%. Candidate \twomu decays are formed from these two muons.
Candidate \twomutwopi decays are formed by combining the trigger muon pair with two oppositely charged
tracks with $\pt>2\GeV$ that do not satisfy the loose muon identification criteria~\cite{CMS:2018rym}, with the pion mass assigned, while candidate \fourmu decays are formed by
combining the trigger muon pair with two additional muon candidates with $\pt>1\GeV$ that satisfy
the loose identification criteria~\cite{CMS:2018rym}. For each channel, the four tracks are required
to have zero total charge and to have a successful fit to a common vertex. The reconstructed \PGh meson candidate is required to have $\pt>10\GeV$ for all three decay channels, as the efficiency for the signal channels is negligible below this value.

Because little energy is released in these decays, assigning the wrong mass hypothesis to two of the tracks shifts the reconstructed invariant mass by about $2(m_{\PGp}-m_{\PGm})\approx70\MeV$. Simulation shows that \twomutwopi decays reconstructed as \fourmu decays are shifted downward by this amount, while \fourmu decays reconstructed as \twomutwopi decays are shifted upward by up to 100\MeV.
The fit range, $0.515 < \meta < 0.595\GeV$, where \meta is the reconstructed four-body invariant mass (\mtwomutwopi or \mfourmu), is chosen such that these misreconstructed contributions lie outside it.

The fraction of events with more than one candidate is negligible in the \fourmu and \twomu channels. 
In the \twomutwopi channel, events are required to have only one candidate within the fit range, which occurs approximately 91\% of the time. 
The efficiency for this selection is obtained from the events simulated with \PYTHIA and used in the branching fraction calculation. 
The robustness of this choice is evaluated using alternative candidate-selection procedures, including retaining all candidates, randomly selecting one candidate, and selecting the candidate with the best vertex fit. 
The maximum difference in the efficiency-corrected signal yields between the nominal and
alternative procedures is 8\%, which is included as a systematic uncertainty in the signal efficiency.

\begin{figure}[th]
\centering
\includegraphics[width=\cmsFigWidth]{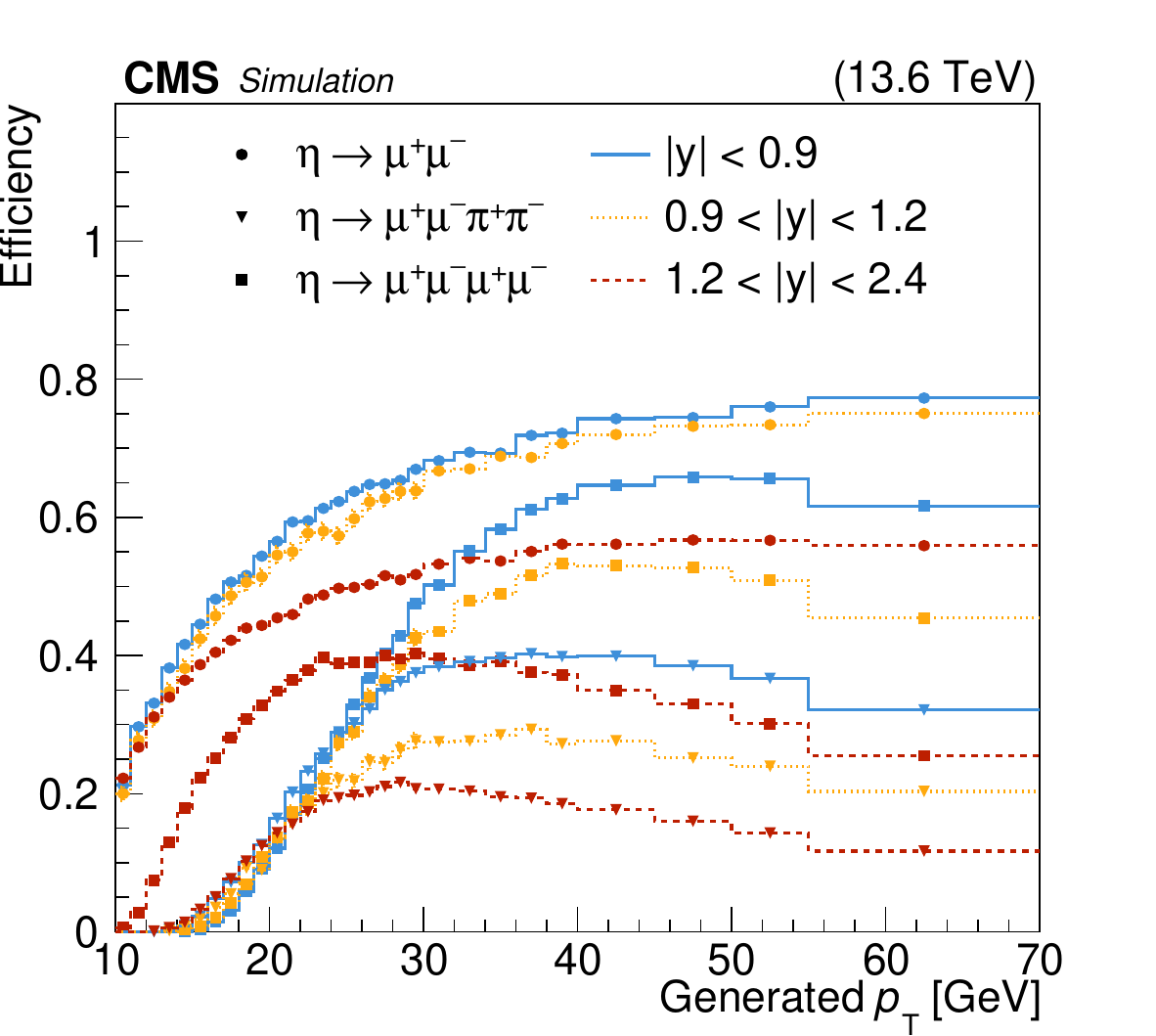}
\caption{Total efficiencies versus \pt of the \PGh meson for the \twomu (circles), \twomutwopi (triangles), and \fourmu (squares) channels, in the three considered $\abs{y}$ ranges: $\abs{y}<0.9$ (blue solid), $0.9<\abs{y}<1.2$ (orange dotted), and $1.2<\abs{y}<2.4$ (red dashed).}
\label{eff}
\end{figure}

The simulated \twomutwopi, \fourmu, and \twomu samples are used to determine the total efficiencies,
$\epsilon_{2\PGm 2\PGp}$, $\epsilon_{4\PGm}$, and $\epsilon_{2\PGm}$, respectively, in 29 bins of \pt and three bins of $\abs{y}$, with \pt bin edges from 10 to 70\GeV and $\abs{y}$ bin boundaries at 0, 0.9, 1.2, and 2.4.
The $\abs{y}$ bin boundaries are chosen to correspond to the barrel, barrel-endcap transition, and endcap regions of the CMS muon system.
Figure~\ref{eff} shows the efficiencies as functions of \pt for different $\abs{y}$ regions.
The efficiency for \fourmu decays is higher than for \twomutwopi decays because the four muons give more chances to satisfy the dimuon trigger requirements. This also explains the high efficiency for \fourmu decays at low \pt in the region $1.2<\abs{y}<2.4$, as muons in this region have higher momentum at the same \pt and are therefore reconstructed more efficiently, and this advantage is preserved by the higher trigger efficiency of the four-muon final state.
The efficiency for \twomutwopi and \fourmu decays decreases with increasing \pt because the Lorentz-boosted \PGh meson produces highly collimated charged particles, making the reconstruction of all four tracks more challenging.

\begin{figure}[th]
\centering
\includegraphics[width=0.49\textwidth]{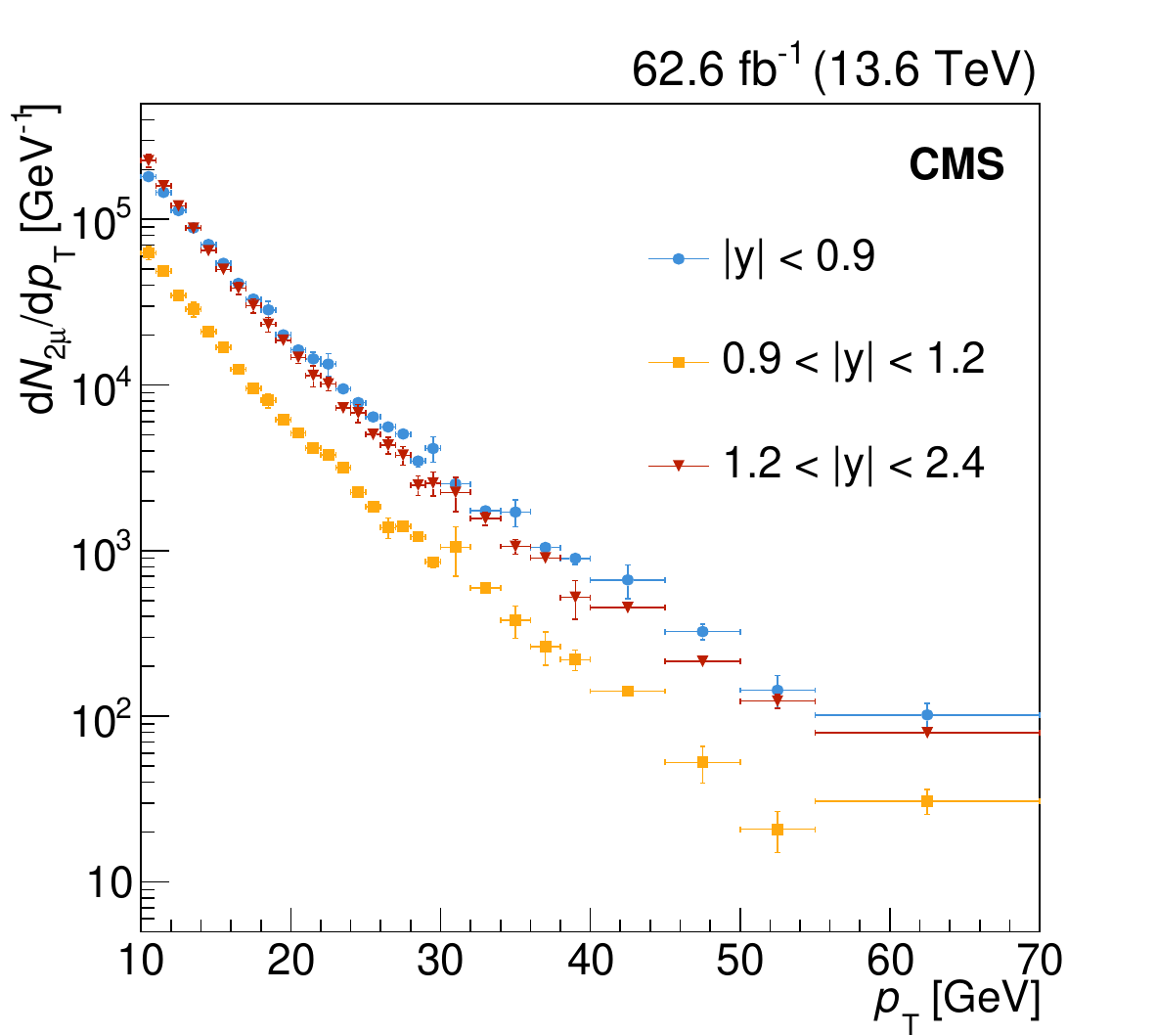}
\caption{Observed \twomu yields as a function of \pt, for the three considered $\abs{y}$ ranges: $\abs{y}<0.9$ (blue circles), $0.9<\abs{y}<1.2$ (orange squares), and $1.2<\abs{y}<2.4$ (red triangles).}
\label{etaproduc}
\end{figure}

The branching fractions of both signal channels are measured relative to the normalization channel \twomu, as shown in Eq.~(\ref{eq:br}). This approach removes the dependence on the \PGh meson production cross section and integrated luminosity.
The \twomu yield is extracted independently in each $(\pt, \abs{y})$ bin by fitting the dimuon invariant mass distribution \mtwomu.
The fit is made with several alternative models, combining single- or
double-Gaussian or Voigtian signal parameterizations with second- or third-order Chebyshev polynomial background functions.
For the double-Gaussian model, the ratio of the two Gaussian normalizations is fixed to the value obtained from simulation, while for the Voigtian model, the Breit--Wigner width parameter is fixed to the simulation value. All other parameters are allowed to float in the fits to data. The mass mean fitted from the measured distribution agrees with the simulation value within 0.15\MeV.
For each $(\pt, \abs{y})$ bin, the yield \Ntwomu is extracted with each fit model.
The nominal yield is taken as the center of the fitted yield range, the statistical uncertainty as the largest of the returned uncertainties, and the systematic uncertainty as the half-range.
The systematic uncertainty is treated as fully correlated across bins, given that changes in the fit model shift the yields coherently.
Figure~\ref{etaproduc} shows the observed \Ntwomu as a function of \pt, in the three considered $\abs{y}$ bins.

The four-body invariant mass signals are modeled with the sum of a reversed Crystal Ball (CB) function~\cite{bib:CB,bib:CB2} and a double-Gaussian function, sharing a common mean.
The shape parameters are determined from simulation separately for the \twomutwopi and \fourmu channels.
Both signal distributions become wider at higher $\abs{y}$, reflecting the degradation of track-\pt resolution. The CB function width increases from 1.6 to 4.1\MeV for \twomutwopi and from 2.1 to 3.9\MeV for \fourmu.
The difference in yield between this function and an alternative parameterization composed of a reversed CB function plus a Voigtian function is assigned as a systematic uncertainty. 
In the fits to data, all signal shape parameters are fixed to the values obtained from simulation, with only the signal normalization left free.

The background in the four-body invariant mass spectrum is modeled using smooth functions that describe the continuum shape.
The nominal choice is a threshold function, \bkgmodela, in which $m_0$ is fixed to the kinematic threshold and $p_0$ is a free parameter. 
Alternative parameterizations, \bkgmodelb and \bkgmodelc, are used to evaluate systematic uncertainties,
where both $p_0$ and $p_1$ are free parameters. The background parameters are left free in the fits to the data.
Potential fit biases are evaluated using ensembles of pseudo-experiments generated from the nominal signal and all background models. For each pseudo-experiment, the fitted signal yield is compared with the injected value, and the pull distribution is tested for consistency with a unit-width Gaussian. No significant bias in the signal yield is observed.

Potential resonant contributions from other processes to both signal channels are evaluated using a simplified simulation
incorporating realistic pion-to-muon misidentification probabilities and \gammatotwomu conversions,
following the framework of previous CMS observations of rare \PGh meson decays~\cite{cms4mu,CMS-PAS-BPH-24-001}.
The reconstructed mass distributions for the \twomugamma, \twopigamma, and \twopipizero decays either lie fully outside the fit range, or extend partially into it.
In the latter case, the contribution is smooth and does not produce a localized peak near the nominal \PGh meson mass. It is therefore absorbed by the background parameterization, resulting in a negligible effect on the extracted signal yield.

Because the observable is constructed as a ratio of signal to normalization yields from two decay processes, many systematic effects partially or fully cancel, as they affect both channels similarly.
Only the residual uncertainties are discussed here and their contributions are incorporated as nuisance parameters in the final likelihood fit.
The uncertainty in the dimuon trigger efficiency does not fully cancel because of kinematic differences between the signal and normalization channels. 
The trigger efficiency is evaluated in the \twomu normalization channel and cross-checked with $\JPsi\to\PGmp\PGmm$ events with the tag-and-probe method~\cite{CMS:2018rym}. The assigned uncertainty is taken as the larger of the observed differences between the two methods and the corresponding statistical uncertainty.
Muon reconstruction and identification efficiencies are measured with the tag-and-probe technique using $\JPsi\to\PGmp\PGmm$ decays~\cite{CMS:2018rym}. The measured differences between the efficiencies in data and simulation are applied to the simulated events as scale factors in bins of muon \pt and $\eta$, and their uncertainties are propagated to the result.
The systematic uncertainty associated with pileup reweighting is evaluated by varying the total
inelastic cross section by $\pm 5\%$~\cite{CMS:2018mlc}. The trigger, muon identification, and pileup uncertainties are propagated to the efficiency ratio $\epsilon_{X}^{i,j}/\epsilon_{2\PGm}^{i,j}$ in each \pt and $\abs{y}$ bin. All such uncertainties are at or below the 1\% level, while those from limited event counts in simulated samples are below 0.1\%.

The uncertainty in the track reconstruction efficiency is nonnegligible because of the differing
track multiplicities of the signal and normalization channels; it is 2.5\% (2.9\%) per track in 2022
(2023), evaluated with the method reported in Ref.~\cite{CMS-DP-2022-012}. Additional systematic uncertainties arise from the candidate selection efficiency and the
choice of signal and background models in the invariant mass fits, including the envelope-based
uncertainty in the extraction of the \twomu yield. 
A summary of the relative systematic uncertainties in the \twomutwopi and \fourmu to \twomu branching-fraction-ratio measurements is given in Table~\ref{tab:systematics}.

\begin{table*}[htbp]
\centering
\topcaption{
Summary of the relative systematic uncertainties used as inputs to the likelihood fit for the \twomutwopi and \fourmu to \twomu branching-fraction-ratio measurements. Asymmetric entries give the relative change in the expected signal yield for the upward and downward variations of each source; the branching fraction ratio changes in the opposite direction.
\label{tab:systematics}}
\begin{scotch}{lcc}
Source & \twomutwopi (\%) & \fourmu (\%) \\
\hline
Trigger efficiency                 & $1.2$ & $0.8$ \\
Muon identification            & $0.3$       & $0.4$ \\
Pileup                                    & $+0.5$/$-0.2$ & $0.1$ \\
Tracking efficiency 2022 (2023)                & $5.0$ ($5.8$)       & $5.0$ ($5.8$) \\
$N_{2\PGm}$ fit model (correlated bins)   & $+8.2$/$-7.3$ & $+7.6$/$-6.9$ \\
$N_{2\PGm 2\PGp}$ or $N_{4\PGm}$ fit model  & $3.0$       & $3.8$ \\
Candidate selection          & $8.0$  & \NA \\
\end{scotch}
\end{table*}

The results are obtained using the CMS statistical analysis tool \textsc{Combine}~\cite{CMS:2024onh}.
Systematic uncertainties are incorporated as nuisance parameters in the likelihood.
The nuisance parameters associated with the trigger efficiency, muon identification efficiency, pileup, and the $N_{2\PGm}$ fit model uncertainties are modeled with Gaussian constraints, while those associated with the tracking efficiency, $N_{2\PGm 2\PGp}$ and $N_{4\PGm}$ fit model uncertainties, and candidate selection efficiency are modeled as log-normal normalization constraints. The former change the expected signal differently in each \pt and $\abs{y}$ bin, while the latter are overall multiplicative factors on the signal yield.

\begin{figure*}[htb]
\centering
\includegraphics[width=0.49\textwidth]{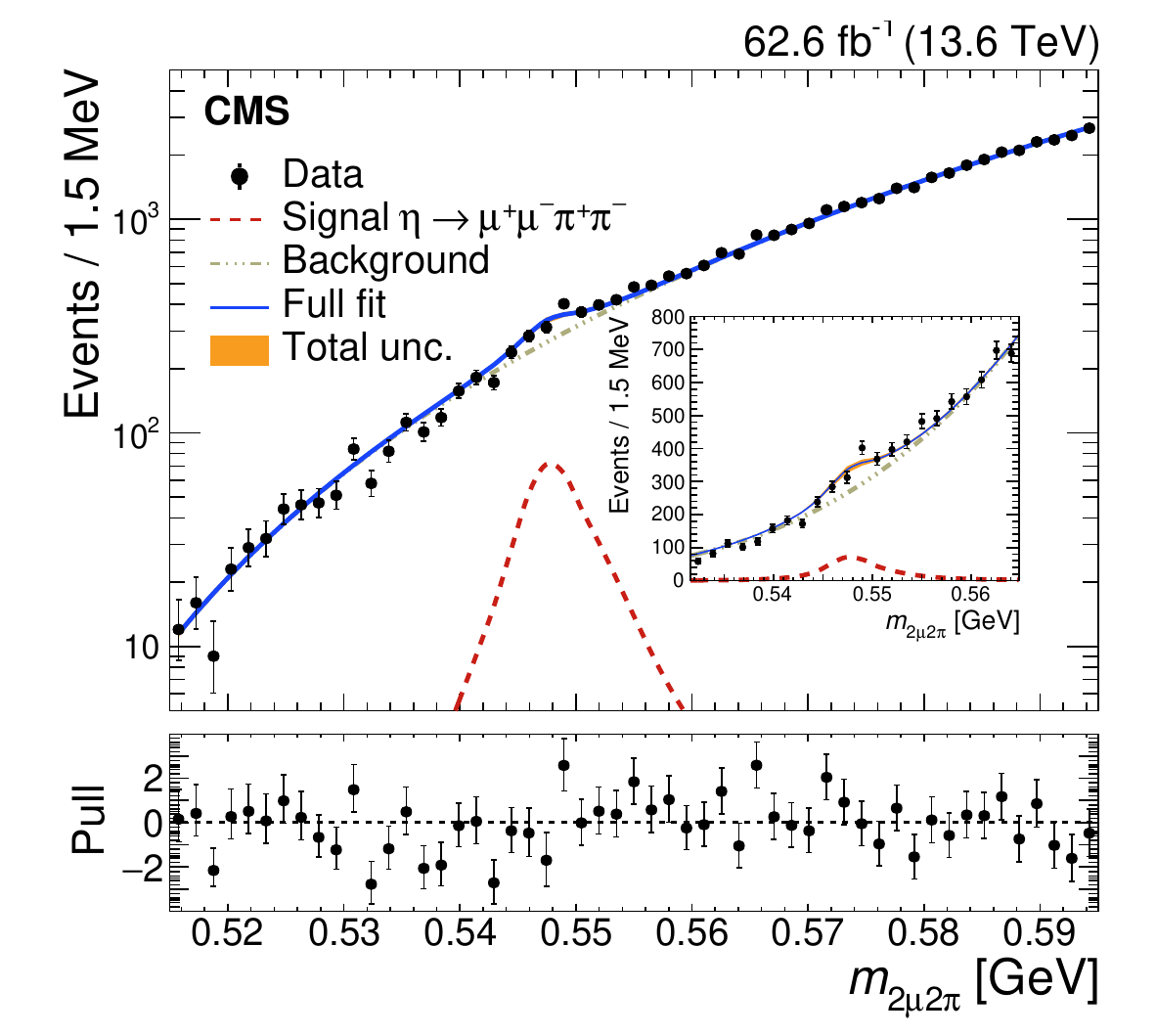}
\includegraphics[width=0.49\textwidth]{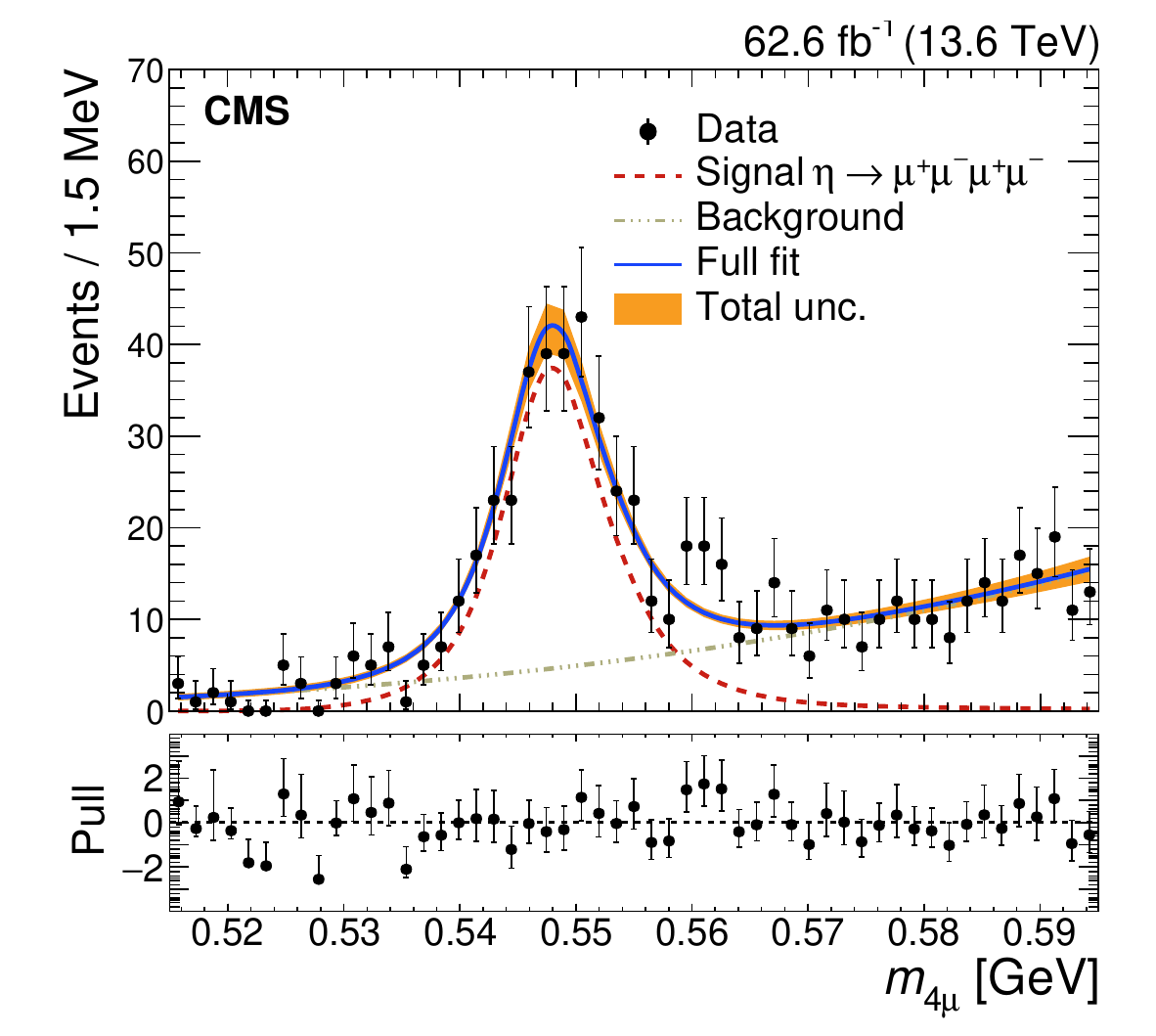}
\caption{
Invariant mass distributions for the \twomutwopi (left) and \fourmu (right) cases, using the nominal background model \bkgmodela.
The inset in the left plot shows the same fit in an expanded region around the \PGh meson mass, with a linear vertical scale.
Black points denote data; the gray curve represents the background component, the red curve the signal component, and the blue curve the total fit (signal + background).
The orange band represents the fit uncertainty, including the statistical and systematic terms.
The lower panel shows the pulls, defined as $(\text{data}-\text{fit})/\sqrt{\smash[b]{\sigma_{\text{data}}^{2}-\sigma_{\text{fit}}^{2}}}$.
}
\label{fit}
\end{figure*}

Figure~\ref{fit} shows the invariant mass distributions for the \twomutwopi and \fourmu candidates in the range 0.515--0.595\GeV.
The parameters of the signal shape are fixed to the simulation prediction, with only the signal normalization allowed to float.
Allowing the signal mean to vary gives a value consistent with the simulation within 0.4\MeV, 
changing the fitted signal yield by about $1\%$ for \twomutwopi and negligibly for \fourmu.
Allowing the signal width to vary gives a
value consistent with the simulation within its uncertainty, changing the 
fitted yield by about 1\% for \twomutwopi and 3\% for \fourmu. 
These shape variations are not included as an additional systematic uncertainty as they are accounted for in the corresponding fit model uncertainties.

The fitted signal yields are $411\pm68$ and $320\pm22$ for the \twomutwopi and \fourmu channels, respectively, where the uncertainties are statistical only.
The signal significance is evaluated using the profile likelihood ratio test statistic
\begin{equation}
q_0 = -2\ln\frac{\mathcal{L}[\mu=0,\hat{\hat{\nu}}(0)]}{\mathcal{L}(\hat{\mu},\hat{\nu})},
\end{equation}
where the nuisance parameters $\nu$ are profiled separately under the
background-only hypothesis ($\mu=0$) and at the best-fit signal strength
$\mu=\hat{\mu}$, assuming the asymptotic approximation~\cite{Cowan:2010js}.
The observed significance, including the effect of all relevant nuisance parameters, is 6.2 standard deviations for \twomutwopi.

The ratios of branching fractions are measured to be
\ifthenelse{\boolean{cms@external}}{
\begin{multline}
\label{br:2mu2pi}
\mathcal{B}(\twomutwopi)/\mathcal{B}(\twomu) = \\ [2.89 \pm 0.48\stat{}^{+0.42}_{-0.29}\syst]\times10^{-3}
\end{multline}
}
{
\begin{equation}
\label{br:2mu2pi}
\mathcal{B}(\twomutwopi)/\mathcal{B}(\twomu) = [2.89 \pm 0.48\stat{}^{+0.42}_{-0.29}\syst]\times10^{-3}
\end{equation}
}
and
\ifthenelse{\boolean{cms@external}}{
\begin{multline}
\label{br:4mu}
\mathcal{B}(\fourmu)/\mathcal{B}(\twomu) = \\ [0.86 \pm 0.06\stat{}^{+0.09}_{-0.08}\syst]\times10^{-3}.
\end{multline}
}
{
\begin{equation}
\label{br:4mu}
\mathcal{B}(\fourmu)/\mathcal{B}(\twomu) = [0.86 \pm 0.06\stat{}^{+0.09}_{-0.08}\syst]\times10^{-3}.
\end{equation}
}
The theoretical predictions of $(6\text{--}12)\times10^{-9}$ for \twomutwopi~\cite{Faessler:1999de,borasoyEtaEtaPrime2007,Escribano_2015,Zillinger:2022eva}
and $(3.98\pm0.15)\times10^{-9}$ for \fourmu~\cite{escribanoDatadrivenApproachPi2018}
are converted into ratios using the weighted average $\mathcal{B}(\twomu)=(5.20\pm0.30)\times10^{-6}$~\cite{BESIII:2025jft,LHCb:2026srw,PDG2026}.
This gives
$\mathcal{B}(\twomutwopi)/\mathcal{B}(\twomu)=(1.2\text{--}2.3)\times10^{-3}$ and
$\mathcal{B}(\fourmu)/\mathcal{B}(\twomu)=(0.77\pm0.05)\times10^{-3}$.
The measured \twomutwopi ratio is about one standard deviation above the upper end of the predicted range, while the measured \fourmu ratio is consistent with the prediction.
The measured \fourmu branching fraction is consistent with the previous CMS measurement~\cite{cms4mu}, with a precision improved by 44\%. 
Combining the two uncorrelated measurements gives
\ifthenelse{\boolean{cms@external}}{
\begin{multline}
\mathcal{B}(\fourmu)/\mathcal{B}(\twomu)=\\(0.86\pm0.09)\times10^{-3}.
\end{multline}
}
{
\begin{equation}
\mathcal{B}(\fourmu)/\mathcal{B}(\twomu)=(0.86\pm0.09)\times10^{-3}.
\end{equation}
}

{\setlength{\emergencystretch}{1em}
In summary, the \twomutwopi decay has been observed, for the first time, on the basis of a sample of proton-proton collisions at $\sqrt{s} = 13.6\TeV$, collected in 2022 and 2023 by the CMS experiment, and corresponding to an integrated luminosity of 62.6\fbinv.
The \twomutwopi and \fourmu branching fractions relative to the \twomu decay are measured with the best precision to date:
$[2.89 \pm 0.48\stat{}^{+0.42}_{-0.29}\syst]\times10^{-3}$ and
$[0.86 \pm 0.06\stat{}^{+0.09}_{-0.08}\syst]\times10^{-3}$, respectively.
Both results are consistent with theoretical predictions.
These results provide new experimental constraints on pseudoscalar transition form factors and precision studies of hadronic effects relevant to the anomalous magnetic moment of the muon.
They also establish benchmarks for future studies of rare \PGh meson decays and for searches for low-mass mediators that could modify these decay rates.
\par}

\begin{acknowledgments}
We congratulate our colleagues in the CERN accelerator departments for the excellent performance of the LHC and thank the technical and administrative staffs at CERN and at other CMS institutes for their contributions to the success of the CMS effort. In addition, we gratefully acknowledge the computing centers and personnel of the Worldwide LHC Computing Grid and other centers for delivering so effectively the computing infrastructure essential to our analyses. Finally, we acknowledge the enduring support for the construction and operation of the LHC, the CMS detector, and the supporting computing infrastructure provided by the following funding agencies: SC (Armenia), BMFWF and FWF (Austria); FNRS and FWO (Belgium); CNPq, CAPES, FAPERJ, FAPERGS, and FAPESP (Brazil); MES and BNSF (Bulgaria); CERN; CAS, MoST, and NSFC (China); MINCIENCIAS (Colombia); MSES and CSF (Croatia); RIF (Cyprus); SENESCYT (Ecuador); ERC PRG and PSG, TARISTU24-TK10 and MoER TK202 (Estonia); Academy of Finland, MEC, and HIP (Finland); CEA and CNRS/IN2P3 (France); SRNSF (Georgia); BMFTR, DFG, and HGF (Germany); GSRI (Greece); MATE and NKFIH (Hungary); DAE and DST (India); IPM (Iran); SFI (Ireland); INFN (Italy); MSIT and NRF (Republic of Korea); MES (Latvia); LMTLT (Lithuania); MOE and UM (Malaysia); BUAP, CINVESTAV, CONACYT, LNS, SEP, and UASLP-FAI (Mexico); MOS (Montenegro); MBIE (New Zealand); PAEC (Pakistan); MSHE, NSC, and NAWA (Poland); FCT (Portugal);  MESTD (Serbia); MICIU/AEI and PCTI (Spain); MOSTR (Sri Lanka); Swiss Funding Agencies (Switzerland); NSTC (Taiwan); MHESI (Thailand); TUBITAK and TENMAK (T\"{u}rkiye); NASU (Ukraine); STFC (United Kingdom); DOE and NSF (USA).
\end{acknowledgments}\section*{Data availability} Release and preservation of data used by the CMS Collaboration as the basis for publications is guided by the  \href{https://doi.org/10.7483/OPENDATA.CMS.1BNU.8V1W}{CMS data preservation, re-use and open access policy}.

\bibliography{auto_generated}
\cleardoublepage \appendix\section{The CMS Collaboration \label{app:collab}}\begin{sloppypar}\hyphenpenalty=5000\widowpenalty=500\clubpenalty=5000\cmsinstitute{Yerevan Physics Institute, Yerevan, Armenia}
{\tolerance=6000
A.~Gevorgyan\cmsorcid{0000-0003-2751-9489}, A.~Hayrapetyan, V.~Makarenko\cmsorcid{0000-0002-8406-8605}, A.~Tumasyan\cmsAuthorMark{1}\cmsorcid{0009-0000-0684-6742}
\par}
\cmsinstitute{Marietta Blau Institute for Particle Physics, Vienna, Austria}
{\tolerance=6000
W.~Adam\cmsorcid{0000-0001-9099-4341}, L.~Benato\cmsorcid{0000-0001-5135-7489}, T.~Bergauer\cmsorcid{0000-0002-5786-0293}, M.~Dragicevic\cmsorcid{0000-0003-1967-6783}, S.~Gundacker\cmsorcid{0000-0003-2087-3266}, A.K.~Guven\cmsorcid{0009-0004-5670-5138}, M.~Jeitler\cmsAuthorMark{2}\cmsorcid{0000-0002-5141-9560}, N.~Krammer\cmsorcid{0000-0002-0548-0985}, A.~Li\cmsorcid{0000-0002-4547-116X}, D.~Liko\cmsorcid{0000-0002-3380-473X}, M.~Matthewman, A.~Pfeiffer\cmsorcid{0000-0001-5328-448X}, J.~Schieck\cmsAuthorMark{2}\cmsorcid{0000-0002-1058-8093}, R.~Sch\"{o}fbeck\cmsAuthorMark{2}\cmsorcid{0000-0002-2332-8784}, M.~Shooshtari\cmsorcid{0009-0004-8882-4887}, N.~Van~Den~Bossche\cmsorcid{0000-0003-2973-4991}, W.~Waltenberger\cmsorcid{0000-0002-6215-7228}, C.E.~Wulz\cmsAuthorMark{2}\cmsorcid{0000-0001-9226-5812}
% ...
\par}
$^{1}$Also at Yerevan State University, Yerevan, Armenia\\
$^{2}$Also at Technische Universit\"{a}t Wien, Vienna, Austria\\
% ...
\end{sloppypar}
\end{document}